\documentclass[12pt]{iopart}

\usepackage{graphicx}
\begin{document}

\title[Image-charge detection of the Rydberg transition of electrons on liquid helium]{Image-charge detection of the Rydberg transition of electrons on superfluid helium confined in a microchannel structure}

\author{S Zou, D Konstantinov}

\address{Quantum Dynamics Unit, Okinawa Institute of Science and Technology (OIST) Graduate University, Okinawa 904-0495, Japan}
\ead{denis@oist.jp}
\vspace{10pt}
\begin{indented} 
\item[]June 2022
\end{indented}

\begin{abstract}
The image-charge detection provides a new direct method for the detection of the Rydberg transition in electrons trapped on the surface of liquid helium. The interest in this method is motivated by the possibility to accomplish the spin state readout for a single trapped electron, thus opening a new pathway towards using electron spins on liquid helium for quantum computing. Here, we report on the image-charge detection of the Rydberg transition in a many-electron system confined in an array of 20-$\mu$m wide and 4-$\mu$m deep channels filled with superfluid helium. Such detection is made possible because of a significant enhancement of the image-charge signal due to close proximity of trapped electrons to the electrodes embedded in the microchannel structure. The transition frequency of electrons in the range of 400-500~GHz is highly controllable by the dc bias voltages applied to the device and is in a good agreement with our calculations. This work demonstrates that microchannel structures provide a suitable platform for electron manipulation and their quantum state detection, with a feasibility of scaling the detection method to a single electron.             
\end{abstract}

\vspace{2pc}

\section{Introduction}
\label{sec:intro}

Electrons bound to the surface of liquid helium present an extremely clean quantum system. The bound states of such surface electrons (SE) are formed due to a weak attractive force from the image charge inside the dielectric liquid and a strong repulsive barrier at the vapor-liquid interface. In such an image potential, the quantized (Rydberg) states of the electron motion perpendicular to the surface have a hydrogen-like spectrum, with an effective Rydberg constant on the order of 1~meV, while the motion parallel to the surface is free. The system is free of disorder and impurities~\cite{Monarkha-book}. At sufficiently low temperatures below 1~K, the scattering of electrons is predominantly from the thermal oscillations (ripplons) of the liquid helium surface~\cite{ShikJLTP1974}. This leads to extremely high mobility of SE and relatively long coherence time of their motional states~\cite{PlatSci1999,DykmPRB2003}. Moreover, the very weak intrinsic spin-orbit (Rashba) interaction and the magnetically-clean environment of superfluid helium-4 result in extremely long coherence time of the spin states of SE~\cite{LyonPRA2004}. This presents SE on superfluid helium-4 as a very attractive system for realizing qubits for quantum computing, with several theoretical proposals presented over the past years~\cite{PlatSci1999,DykmPRB2003,LyonPRA2004,SchuPRL2010}.

While unique transport properties of a many-electron system on liquid helium had been extensively investigated in the past~\cite{Andr-book}, methods for the quantum state control of SE, in particular at the level of a single electron, were not sufficiently developed. The lack of such methods presented the main obstacle towards the realization of the above proposals for qubit implementation. Very recently, a readout of the quantized states of the trapped lateral motion was demonstrated for a single electron on superfluid helium~\cite{KostNatComm2019} and on solid neon~\cite{ZhouNat2022}. In these experiments, the electrons were confined in an electrostatic trap integrated into a circuit quantum electrodynamic (cQED) architecture~\cite{BlaiRMP2021}. The above works present the first significant advance towards qubit implementation using SE, with a further promise to realize coupling between the motional and spin states of an electron, thus a readout of its spin states~\cite{SchuPRL2010}. 

An alternative approach for the spin-state readout was recently suggested by Kawakami {\it et al.} who proposed to use a coupling between the spin states of an electron and the Rydberg states of its vertical motion~\cite{KawaPRL2019}. The idea of readout is based on a relatively large difference between the spatial extension of the electron wavefunction above the liquid surface for an electron in the ground state and the first excited Rydberg state, respectively. Thus, in the presence of a sufficiently large gradient of an applied magnetic field, for example created by a magnet placed in proximity to SE, an electron in different Rydberg states will experience different values of the mean magnetic field, resulting in distinct Zeeman splitting of their spin states. Consequently, the transition frequency between the ground state and the first excited Rydberg state of an electron will depend on its spin state. Then, the Rydberg transition can be excited spin-selectively by an applied resonant microwave (MW) radiation providing that the above difference in the Zeeman splitting exceeds the Rydberg transition linewidth. In other words, the spin state of an electron can be determined by detecting its Rydberg transition.

For this purpose, a new method of direct detection of the Rydberg transition in SE, the image-charge detection, was developed~\cite{KawaPRL2019}. The idea of this method is also based on the difference in the spatial extension of the electron wavefunction above the surface for different Rydberg states. A change in the Rydberg state of an electron causes a corresponding change in the image charges induced by the electron in surrounding metal electrodes. Thus, the Rydberg transition in SE can be detected by measuring the image current at one of the electrodes. The proposed method was successfully demonstrated by placing a large number of SE in the middle between two plates of a parallel-plate capacitor separated by a distance $D=2$~mm and measuring the image current induced in the plates by SE in response to a pulse-modulated MW excitation at the modulation frequency $\sim$100~kHz. For liquid helium-4, the mean distance between an electron and the liquid surface is equal to 11.4 and 45.6~nm for the electron occupying the ground state and the first excited Rydberg state, respectively, with the corresponding difference $\Delta z\approx 34$~nm~\cite{Monarkha-book}. Thus, each excited electron causes a change in the image charge at the bottom and top plates of the capacitor equal to $\delta q=\mp e\Delta z/D$, respectively, where $e>0$ is the elementary charge. This amounts to $|\delta q|\sim 10^{-5}e$ for the setup used in \cite{KawaPRL2019}. For about $10^8$ electrons used in the experiment, this produced a current on the order 10~pA, which was directly measured using a lock-in amplifier. 

The ultimate goal is the image-charge detection of the Rydberg transition of a single electron. The simplest improvement towards increasing the sensitivity of the proposed method is immediately seen, that is increasing the change in the image charge $|\delta q|$. As follows from the above discussion, this can be done by decreasing the distance $D$ between the capacitor plates. Unfortunately, for a parallel-plate capacitor partially filled with liquid helium this distance can not be made less than $\sim 1$~mm, the capillary length of liquid helium, otherwise the capacitance will be completely filled with the liquid by the capillary action~\cite{LeaPRB2017}. For this reason, other electrode structures have to be sought, for example, a coplanar capacitor covered with a superfluid helium film, as was suggested in \cite{KawaPRL2019}. An alternative approach is to use the so called microchannel devices, which have been very useful to study the transport properties of SE on superfluid helium~\cite{GlasPRL2001,IkegPRL2009,ReesPRL2016,BadrPRL2020}. Such a device consists of an array of microchannels, with a typical channel depth $d\sim 1$~$\mu$m, which can be filled with the superfluid helium-4 by the capillary action. SE floating on the surface of superfluid helium inside a channel are capacitively coupled to the metal electrodes comprising the top and bottom of the channel, therefore they are expected to produce a change in the image charge induced at one of the electrodes equal to $|\delta q | = \kappa e\Delta z/d\sim 10^{-2}e$, where $\kappa$ is a numerical factor order of unity, which is determined by the ratio of the capacitive coupling of SE to each of the two channel electrodes. Thus, the sensitivity of the image-charge detection method is expected to enhance thousandfold for SE in a microchannel device comparing with the previous experiment~\cite{KawaPRL2019}. In addition, the microchanel devices present several other advantages towards spin-state control and quantum computing. The micromagnets~\cite{TokuPRL2006}, which are required to introduce a sufficiently strong magnetic field gradient, can be easily integrated to the microchannel structure, therefore can be placed in close proximity to SE. It was also demonstrated that, for the sake of increasing functionality of a device, the microchannel geometry can be easily integrated with other useful mesoscopic devices, for example, the single-electron transistor (SET)~\cite{PapaAPL2005,RousPRB2009}, point-contact constriction~\cite{ReesPRL2011}, superconducting coplanar waveguide (CPW) resonator~\cite{KostNatComm2019,ZhouNat2022}, and charge-coupled device (CCD)~\cite{BradPRL2011}. The latter is particularly attractive towards realization of a quantum-CCD architecture for scalable quantum computing~\cite{KielNat2002,YuPRA2022}. Finally, the Rydberg resonance of SE confined in a single 4-$\mu$m deep channel was recently observed by an indirect resistive detection method, thus showing feasibility of the Rydberg state detection in such a system~\cite{ZouJLTP2022}.  

Here, we report on the first image-charge detection of the Rydberg transition in SE on superfluid helium confined in a microchannel structure consisting of an array of 20-$\mu$m wide and 4-$\mu$m deep microchannels. An image-current signal from approximately $10^5$ electrons, whose Rydberg resonance is excited by a pulse-modulated MW radiation in the frequency range 400-500~GHz, is converted to voltage by a two-stage cryogenic amplifier and recorded at room temperature by a lock-in amplifier. The microchannel device used in our experiment shows high-level control of the transition frequency of confined SE via the Stark shift of the Rydberg energy levels. Further improvement in the image-current detection circuit promises the feasibility to enhance sensitivity of our method to the detection of a single electron. Thus, our work demonstrates that SE on superfluid helium confined in a microchanel device present a suitable platform for realizing the quantum state detection in electrons on helium for their qubit implementation.                           

\section{Method}
\label{sec:method}

\subsection{Image-charge detection}
\label{subsec:image}

\begin{figure}[t]
\begin{center}
\includegraphics[width=16cm,keepaspectratio]{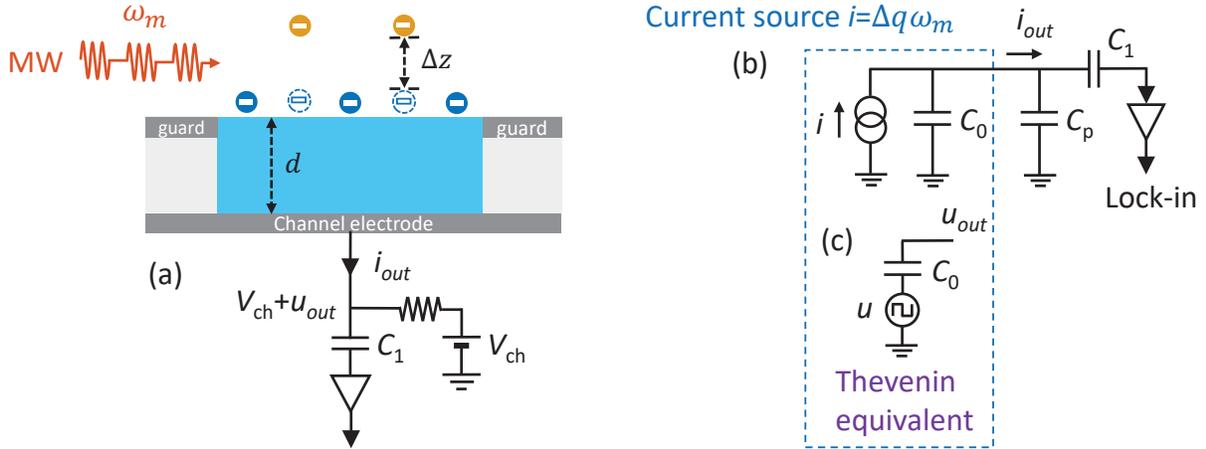}
\end{center}
\caption{\label{fig:1} Schematic view of the image-charge detection method (a) and its equivalent Norton and Thevenin electric circuits, (b) and (c), respectively. The image current $i_{out}$ at the channel electrode is generated by the pulse-modulated MW excitation of SE in the channel with the modulation frequency $\omega_m$.}
\end{figure}

A schematic view of the image-charge detection method, as well as its equivalent electric circuit, for SE on superfluid helium confined in a channel are shown in figure~\ref{fig:1}. Figure~\ref{fig:1}(a) schematically shows the cross-section of a channel of depth $d$ filled with superfluid helium. Electrons trapped on the surface of helium are capacitively coupled to the channel and guard electrodes  at the bottom and top of the channel, respectively. A positive dc bias voltage $V_{\rm ch}$ is applied to the channel electrode to create an electrostatic confining potential for SE. For the sake of simplicity, we assume that the guard electrode is grounded. Excitation of SE by the pulse-modulated MW radiation with the modulation frequency $\omega_{\rm m}$ causes an ac current $i_{\rm out}$ in the channel electrode due to the periodic variation, also with the frequency $\omega_{\rm m}$, of the electric charge induced in the electrode. This current flows to the detection circuit via a coupling capacitance $C_1$, which forms a bias-T with a resistor at the dc voltage source. Thus, the voltage at the channel electrode is given by $V_{\rm ch}+u_{\rm out}$, where $u_{\rm out}$ is the ac voltage induced by $i_{\rm out}$. 

The total electric charge at the channel electrode is given by $(Q+q)$, where $Q=C_0(V_{\rm ch}+u_{\rm out})$ and $C_0$ is the capacitance between the channel and guard electrodes determined by the channel geometry, while $q$ is the image charge induced by SE. Thus, the output current is given by 

\begin{equation}
i_{\rm out}=-\frac{d(Q+q)}{dt} = - C_0 \frac{d u_{\rm out}}{d t} - \frac{d q}{d t}.
\label{eq:1}
\end{equation} 

\noindent The variation of the image charge at the channel electrode due to excitation of SE can be estimated as $\Delta q =  \rho_{22}  N_{\rm e} \delta q = - \rho_{22} N_{\rm e} \kappa e\Delta z/d$, where $N_{\rm e}$ is the total number of SE in the channel and $\rho_{22}$ is the fraction of these electrons that is excited by the MW radiation. For the microchannel geometry used in our experiment, we estimate $\kappa = 0.8$~\cite{ZouJLTP2022}. For the pulse-modulated MW excitation, the corresponding current is given by $i=-(dq/dt)=-\Delta q \omega_{\rm m}$. Thus, as follows from (\ref{eq:1}), an equivalent electric circuit of our setup can be represented as the Norton circuit with an ideal current source $i$, see figure~\ref{fig:1}(b). Here, in addition to the coupling capacitance $C_1$, we included a parasitic capacitance $C_{\rm p}$ which in our experiment is mostly determined by the capacitance of a cryogenic cable connecting the microchanel device to the detection circuit. 

\subsection{Detection circuit}

In the experiment reported here, we used a two-stage cryogenic amplifier that detected the voltage signal $u_{\rm out}$ due to the image current $i_{\rm out}$, with a subsequent detection using a room-temperature lock-in amplifier referenced at the modulation frequency $\omega_{\rm m}$. A detailed description of the amplifier circuit is given elsewhere~\cite{ElarJLTP2021}. The first stage of the amplifier was located in close proximity to the experimental cell, with an estimated parasitic capacitance of the connecting cable $C_{\rm p}\approx 20$~pF. The two-stage amplifier provided a total voltage gain of $\sim 100$.

For the sake of comparison between our theoretical estimations and experimental results, it is convenient to transform the Norton circuit given in figure~\ref{fig:1}(b) to its Thevenin equivalent shown in figure~\ref{fig:1}(c), with an ideal voltage source $u=i/(\omega_{\rm m} C_0)=-\Delta q/C_0$. Assuming that the first stage of the cryogenic amplifier has an infinitely large input impedance~\cite{ElarJLTP2021}, the voltage signal $u_{\rm out}$ at the input of the amplifier is given by $u_{\rm out}=uC_0/(C_0+C_{\rm p})=-\Delta q/(C_0+C_{\rm p})$. Note that the same result can be also obtained from (\ref{eq:1}). Thus, we obtain

\begin{equation}
u_{\rm out}=\frac{\rho_{22} N_{\rm e} \kappa e\Delta z}{d(C_0+C_{\rm p})}.
\label{eq:sign}
\end{equation}

\noindent To estimate the magnitude of the expected measured signal, we take $\Delta z/d=10^{-2}$, $N_{\rm e}=10^5$, $\rho_{22}=0.1$ (10\% population of the excited Rydberg state), and $C_0=1.3$~pF, which is estimated from the device geometry, thus obtaining $u_{\rm out}~\approx 5$~$\mu$V. This corresponds to an expected lock-in output signal of 0.5~mV.       
       
\subsection{Microchannel device}

\begin{figure}[t]
\begin{center}
\includegraphics[width=16cm,keepaspectratio]{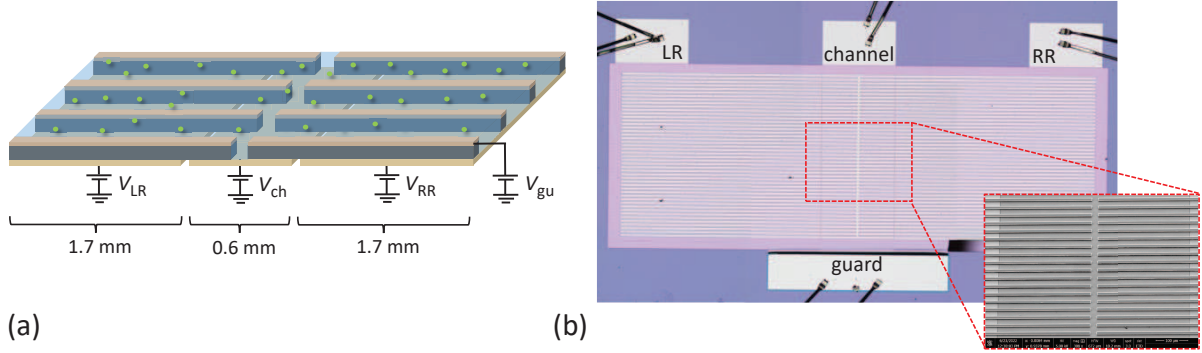}
\end{center}
\caption{\label{fig:2} Schematic view of the microchannel device (a) and its optical microscope image showing the bonding pads (b). The inset is SEM micrograph zooming the middle segment of the microchannel array.}
\end{figure}

A schematic view of the microchanal device used in our experiment, as well as its optical and SEM micrographs, are shown in figure~\ref{fig:2}. The device consists of an arrays of 47 connected channels, each 20-$\mu$m wide, arranged in parallel. The whole structure is fabricated on a silicon substrate using UV lithography methods and is composed of two thin patterned gold layers separated by an insulating layer of hard-baked photoresist with a thickness $d=4$~$\mu$m, which defines the depth of the channels. The channel array is divided into three segments, each having a length of 1.7, 0.6, and 1.7~mm, respectively. The decomposition into segments is defined by the structure of the bottom layer, which is divided into three electrodes separated by 2-$\mu$m wide gaps to form bottom electrodes for each of the three segments. The corresponding dc bias voltages applied to these electrodes are denoted as $V_{\rm LR}$, $V_{\rm RR}$, and $V_{\rm ch}$, see figure~\ref{fig:2}(a). The two 1.7-mm long segments serve as left and right electron reservoirs (LR and RR, respectively) which are connected by the third 0.6-mm long middle segment. The top gold layer consists of a single electrode, the guard electrode, which lies in the plane of the liquid helium surface. The guard electrode defines the channel boundary and, together with the bottom electrodes, serves to provide the confinement of SE inside the channels and their tuning to the Rydberg resonance via the Stark shift, as will be discussed in section~\ref{sec:exp}. The corresponding dc bias voltage applied to this electrode is denoted as $V_{\rm gu}$.   

\subsection{Device operation}

The microchannel device is mounted on a printed circuit board (PCB) and bonded with the conducting pads of PCB using a 50-$\mu$m diameter gold wire, see figure~\ref{fig:2}(b). PCB is placed inside a leak-tight copper cell and is connected to the control and detection circuits via hermetic SMP connectors mounted at the top flange of the cell. The cell is cooled down below 1~K at the mixing chamber of a dilution refrigerator. Helium-4 gas is condensed into the cell, such that the liquid level is placed slightly below PCB, and superfluid helium fills the channels of the device by the capillary action. The surface of liquid in the channels is charged with electrons produced by the thermionic emission from a tungsten filament placed about 1~mm above the device, with the applied bias voltages $V_{\rm LR}=V_{\rm RR}=V_{\rm ch}=1$~V and $V_{\rm gu}=0$~V. 

The Rydberg transition of SE in the microchannel device is excited by the pulse-modulated MW radiation, which is supplied from a room-temperature source, guided into the cryostat through an overmoded (WR-28) stainless steel waveguide and sent into the experimental cell through a tapper waveguide transition. The MW power measured at the output of the source was about 1~mW in the frequency range 350-500~GHz, while the total transmission loss of the waveguide system was measured to be about -10~dB. The MW power at the location of the microchannel device inside the cell could not be directly measured but is estimated to be in the range $10\textrm{--}100$~$\mu$W.   

In the experiment reported here, SE are excited by the MW radiation at a fixed frequency $\omega$, while the  resonant frequency $\omega_{21}$ corresponding to the transition of SE from the ground state to the first excited Rydberg state is tuned in resonance with $\omega$ via the Stark shift by varying the pressing electric field $E_\perp$ exerted on the electrons perpendicular to the surface of liquid~\cite{LeaPRL2002}. This field is determined by the dc bias voltages applied to the device electrodes, as well as by the image charges induced in the electrodes by SE. The latter depend on the density of SE in the channel and the channel depth. The image current due to the MW-excited electrons is measured at the bottom electrode of the middle segment, while the bias voltages applied to the device electrodes are varied to change $E_\perp$. Note that a relatively large depth of 4~$\mu$m allows us to apply relatively large electrode voltages, thus to preserve the resolution of the transition line while voltages are swept through the resonance. For typical density of SE in our device~\cite{ZouJLTP2022} and the total middle-segment channel area of about $6\times 10^{-3}$~cm$^{-2}$, we expect to have on the order $10^5$ electrons in the middle segment.

\section{Results}
\label{sec:exp}

\begin{figure}[t]
\begin{center}
\includegraphics[width=10cm,keepaspectratio]{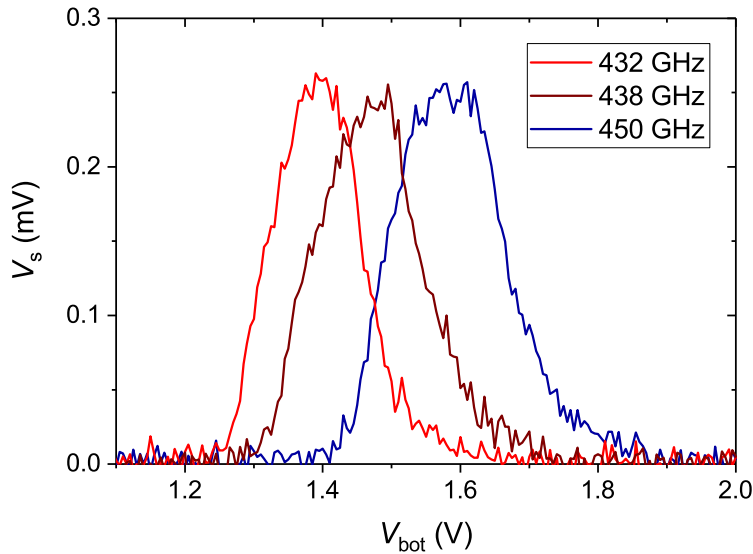}
\end{center}
\caption{In-phase voltage signals recorded by the lock-in amplifier referenced at the modulation frequency $\omega_{\rm m}/2\pi=100$~kHz versus the bottom electrode voltage $V_{\rm bot}=V_{\rm LR}=V_{\rm RR}=V_{\rm ch}$. Each trace is recorded with a fixed value of the MW frequency $\omega/2\pi$ given in the plot legend.}
\label{fig:3}
\end{figure}

\begin{figure}[t]
\begin{center}
\includegraphics[width=16cm,keepaspectratio]{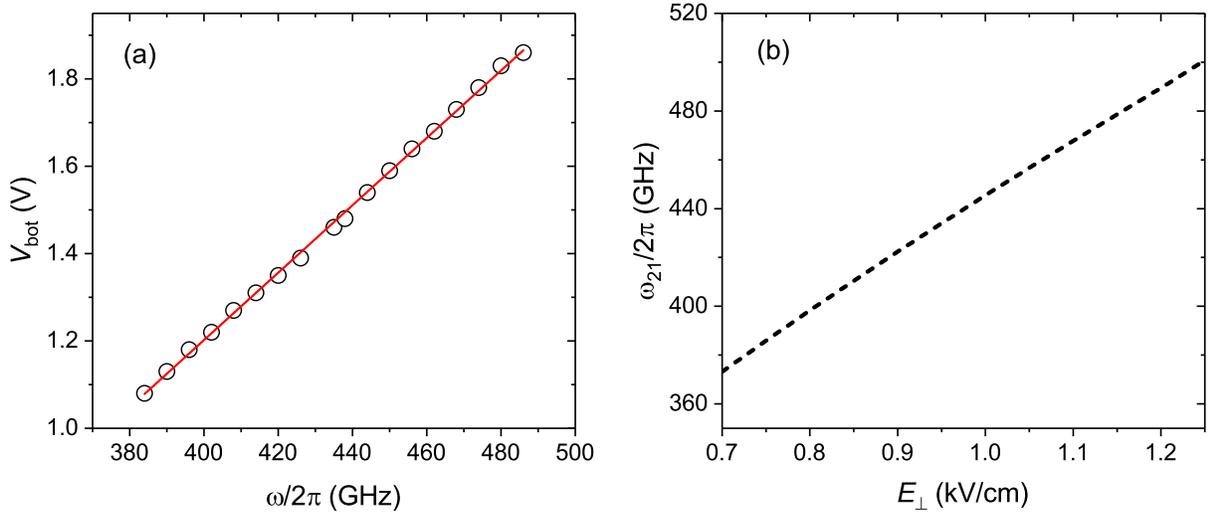}
\end{center}
\caption{(a) Position of the maxima of the resonance signals observed in the experiment versus the MW frequency (symbols). The red line is a linear fit to the experimental data with the slope $k_1=0.0077$~V/GHz. (b) Numerically calculated dependence of the transition frequency $\omega_{21}/2\pi$ of SE on liquid helium-4 for different values of the pressing electric field $E_\perp$.}
\label{fig:4}
\end{figure}

Figure~\ref{fig:3} shows some exemplary image-charge signals from the excited SE for several values of the MW frequency $\omega$ while the Rydberg transition frequency of SE is tuned in resonance by varying the bias voltage $V_{\rm bot}$ applied to the bottom electrodes, with $V_{\rm bot}=V_{\rm LR}=V_{\rm RR}=V_{\rm ch}$ and $V_{\rm gu}=0$. Note that by keeping the voltages applied to all three bottom electrodes equal to each other, we ensure that the density of electrons across all three segments is uniform and constant, by the conservation of the total number of SE in the device. The data reported here were taken at $T=0.2$~K, which was measured by a ruthenium oxide sensor attached to the experimental cell. At such temperature, the thermal population of the excited Rydberg states is negligible. The in-phase voltage signal $V_{\rm s}$ is recorded by a lock-in amplifier referenced at the MW modulation frequency $\omega_{\rm m}/2\pi=100$~kHz and with the time constant of 1~s. The resonant peak for each trace corresponds to the transition of SE from the ground state to excited Rydberg states. The position of resonance shifts towards higher values of $V_{\rm bot}$ with increasing MW frequency, consequence of the linear Stark shift of the Rydberg energy levels of SE~\cite{LeaPRL2002}. Figure~\ref{fig:4}(a) shows a summary plot (symbols) of the position of the observed resonance peaks versus $\omega/2\pi$, which demonstrates the linear dependence with the slope $k_1=0.0077$~V/GHz (red line). Using this result, we can estimate the average half-width at half-maximum (HWHM) of the resonance peaks of about 10~GHz. At $T=0.2$~K, the intrinsic linewidth of the Rydberg resonance due to the scattering of electrons from ripplons is expected to be of the order of 10~MHz~\cite{AndoJPSJ1978}. A thousandfold larger broadening observed in our experiment is due to the inhomogeneous broadening of the Rydberg transition of SE caused by the variation of the pressing electric field $E_\perp$ across the channel. This result is consistent with the numerically calculated distribution of $E_\perp$ for our channel geometry~\cite{ZouJLTP2022}. Note that, comparing to the resonance signal measured by the indirect resistive method~\cite{ZouJLTP2022}, a typical resonance peak reported here is much more symmetric, as should be expected from the electric field distribution in the given channel geometry. This is because the image-charge detection method employed here directly probes the excited-state population $\rho_{22}$ of SE, as follows from (\ref{eq:sign}). Also, note that the observed magnitude of the image-charge resonant signal is in a good agreement with our theoretical estimation given in section~\ref{subsec:image}.  

In order to compare the position of the resonance signal with theory, we numerically solved the eigenvalue problem for the quantized motion of an electron perpendicular to the liquid helium surface to find the dependence of its transition frequency $\omega_{21}$ on the electric field $E_\perp$. The result is shown in figure~\ref{fig:4}(b). In the given range of $E_\perp$, this dependence is almost linear with the slope $k_2 = 0.23$~GHz/(V/cm). Assuming that the resonance signal observed in the experiment corresponds to the resonant condition $\omega=\omega_{21}$, we obtain the linear relation between the pressing electric field $E_\perp$ acting on SE and the applied bias voltage $V_{\rm bot}$, with the slope equal to $k=(k_1 k_2)^{-1}=565$~cm$^{-1}$. In \cite{ZouJLTP2022}, an approximate theoretical relation between $E_\perp$ and the voltages applied to the device electrodes was established, which for our device reads

\begin{equation}
E_\perp = \frac{V_{\rm bot}}{2d} - \frac{\beta V_{\rm gu}}{2\alpha d} - \frac{V_{\rm e}}{d} \left( 1-\frac{1}{2\alpha} \right).
\label{eq:Eperp1}
\end{equation}  

\noindent Here, $\alpha$ and $\beta$ are weighted contributions ($\alpha + \beta =1 $) to the total capacitance between the charged surface of liquid and device electrodes due to the bottom channel electrode and the top guard electrode, respectively, while $V_{\rm e}$ is the electric potential of the charged surface. For our channel geometry, we have $\alpha=0.8$ and $\beta=0.2$~\cite{ZouJLTP2022}. To account for the experimental conditions in figure~\ref{fig:4}(a), we set $V_{\rm gu}=0$. Moreover, unlike for the device with a single-channel middle segment used in \cite{ZouJLTP2022}, for the current device with a multi-channel middle segment the potential $V_{\rm e}$ depends on the value of the applied bias voltage $V_{\rm bot}$. In order to find the relationship between $V_{\rm e}$ and $V_{\rm bot}$, we notice that the electric potential of the uncharged surface of liquid in the channel is given by $V_{\rm u}=\alpha V_{\rm bot} + \beta V_{\rm gu}=\alpha V_{\rm bot}$ (with $V_{\rm gu}=0$). The difference between the electric potentials of the uncharged and charged surface of liquid $\Delta V = V_{\rm u} - V_{\rm e}$ depends only on the number of SE and the channel geometry, therefore is fixed during the variation of $V_{\rm bot}$. Plugging the relation $V_{\rm e}=\alpha V_{\rm bot} - \Delta V$ into (\ref{eq:Eperp1}), we obtain

\begin{equation}
E_\perp = \frac{V_{\rm bot}}{d} (1-\alpha) + \frac{\Delta V}{d} \left( 1-\frac{1}{2\alpha} \right),
\label{eq:Eperp2}
\end{equation} 

\noindent which predicts the linear dependence between $E_\perp$ and $V_{\rm bot}$ with the slope $(1-\alpha)/d=500$~cm$^{-1}$. This is fairly close the the experimentally observed slope of $565$~cm$^{-1}$. Note that in an actual device the charged surface of liquid helium in the channel is somewhat depressed due to the force exerted on SE by the electric field $E_\perp$. The smaller value of $d$ results in the higher value of the slope, in agreement with our result.

\begin{figure}[t]
\begin{center}
\includegraphics[width=10cm,keepaspectratio]{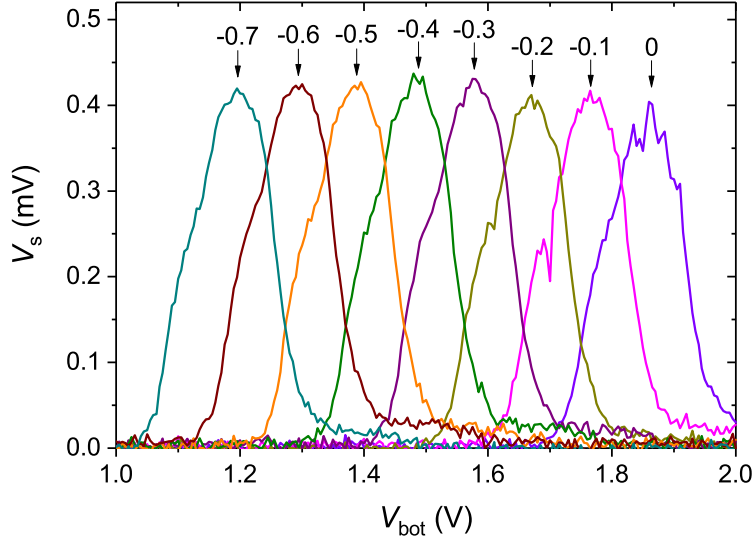}
\end{center}
\caption{In-phase voltage signal recorded by the lock-in amplifier referenced at the modulation frequency $\omega_{\rm m}/2\pi=200$~kHz versus the bottom electrode voltage $V_{\rm bot}=V_{\rm LR}=V_{\rm RR}=V_{\rm ch}$. Each trace is recorded with for the same MW frequency $\omega/2\pi=480$~GHz and different values of the negative dc bias voltage  $V_{\rm gu}$ applied to the guard electrode, which is indicated (in Volts) for each trace.}
\label{fig:5}
\end{figure} 

The above relation (\ref{eq:Eperp1}) also suggests that the resonance position in $V_{\rm bot}$ for a fixed MW frequency $\omega$ can be adjusted by varying the guard voltage $V_{\rm gu}$. Fig.~\ref{fig:5} shows the resonant image-charge signals for MW radiation with a fixed frequency 480~GHz and for several fixed negative biases $V_{\rm gu}$. The position of the resonance shifts towards smaller values of $V_{\rm bot}$ for more negative values of $V_{\rm gu}$. The plotted dependence of $V_{\rm bot}$ at resonance versus $V_{\rm gu}$ is linear with the slope very close to 1. Using the above relationship between $V_e$ and the applied electrode voltages, with $V_e=\alpha V_{\rm bot} + \beta V_{\rm gu} - \Delta V$, we obtain from (\ref{eq:Eperp1})   

\begin{equation}
E_\perp = \frac{V_{\rm bot}}{d} (1-\alpha) - \frac{V_{\rm gu}}{d}\beta - \frac{\Delta V}{d} \left( 1-\frac{1}{2\alpha} \right).
\label{eq:Eperp3}
\end{equation} 

\noindent This equation predicts the linear relationship between the pressing field $E_\perp$ and the voltage difference $V_{\rm bot} - V_{\rm gu}$ with slope equal to $\beta/d$. This is in an excellent agreement with our experimental result shown in figure~\ref{fig:5}, where SE are excited by the MW radiation at a fixed frequency $\omega$ corresponding to a certain value of $E_\perp$, while the values of $V_{\rm bot}$ and $V_{\rm gu}$ satisfy the relation $V_{\rm bot}-V_{\rm gu} \simeq 1.88$~V, with the slope of the linear relationship between $V_{\rm bot}$ and $V_{\rm gu}$ equal to 1.  

\begin{figure}[t]
\begin{center}
\includegraphics[width=16cm,keepaspectratio]{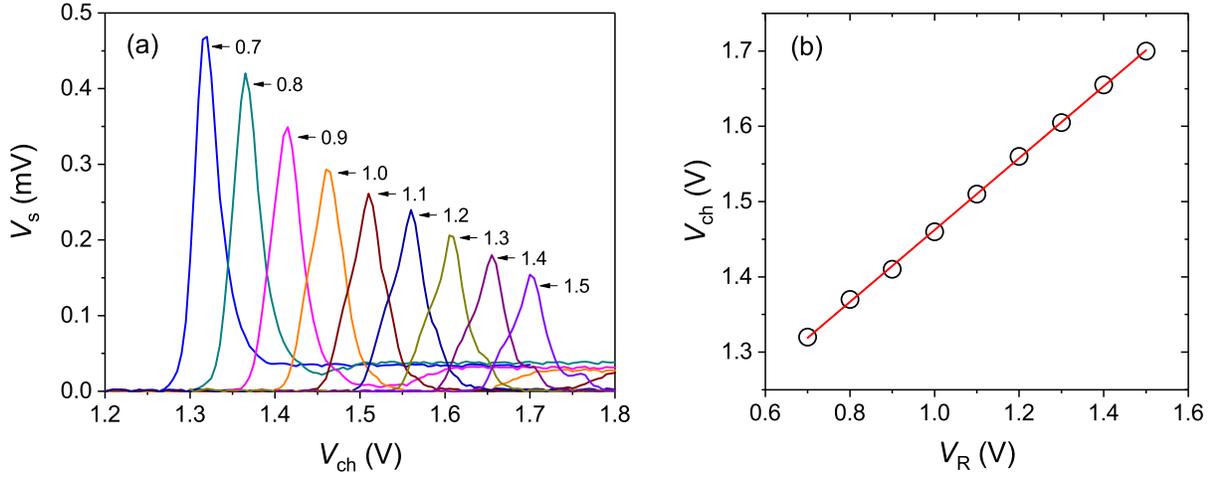}
\end{center}
\caption{(a) In-phase voltage signal recorded by the lock-in amplifier referenced at the modulation frequency $\omega_{\rm m}/2\pi=200$~kHz versus the channel electrode voltage $V_{\rm ch}$. Each trace is recorded for the same MW frequency $\omega/2\pi=480$~GHz and different values of the positive dc bias voltage $V_{\rm R}=V_{\rm LR}=V_{\rm RR}$ applied to the reservoir electrodes, as indicated (in Volts) for each trace. (b) Position of the maxima of the resonance signals shown in (a) versus $V_{\rm R}$ (symbols). The red line is a linear fit to the experimental data with the slope equal to approximately $0.48$.}
\label{fig:6}
\end{figure} 

As mentioned earlier, in the experiments described above the bias voltages $V_{\rm ch}$, $V_{\rm LR}$ and $V_{\rm RR}$ applied to all three bottom electrodes of the device were varied simultaneously to ensure that the density of SE in the middle segment is constant during the voltage sweep. Alternatively, we can fix the dc bias $V_{\rm R}$ applied to the reservoir electrodes and tune the transition frequency of SE in the middle segment into resonance with MWs by varying $V_{\rm ch}$. In this case, the density of SE in the middle segment is expected to vary as electrons redistribute between the middle segment and reservoirs, while keeping the total number of SE fixed. Figure~\ref{fig:6}(a) shows the resonant image-charge signals for the MW radiation with a fixed frequency 480 GHz and for several fixed positive biases $V_{\rm R}$, with $V_{\rm R}=V_{\rm LR}=V_{\rm RR}$ and $V_{\rm gu}=0$. The resonance shifts towards higher values of $V_{\rm ch}$ with increasing $V_{\rm R}$, while the signal amplitude decreases. Qualitatively, both effects appear because the density of electrons in the middle segment decreases with increasing $V_{\rm R}$. In figure~\ref{fig:6}(b), we plot the position of the resonance peaks shown in figure~\ref{fig:6}(a) versus the reservoir voltage $V_{\rm R}$. The dependence is linear (red line) with the slop equal to approximately 0.48.   

In order to quantitatively account for the observed resonance shift, we note that the density of SE in the middle segment varies with the bias voltages applied to the device electrodes according to~\cite{ZouJLTP2022}

\begin{equation}
n_s=\frac{\varepsilon\varepsilon_0}{\alpha e d} \left( \alpha V_{\rm ch} + \beta V_{\rm gu} - V_{\rm e}\right),
\label{eq:ns}
\end{equation}

\noindent where $\varepsilon$ is the dielectric constant of liquid helium and $\varepsilon_0$ is the vacuum permittivity. A similar equation is valid for the density of SE in the reservoirs, with $V_{\rm ch}$ replaced by $V_{\rm R}$, owing to the fact that the electric potential of the charged surface of liquid $V_{\rm e}$ is uniform across the whole device. Thus, we can establish the following relationship between $V_{\rm e}$, $V_{\rm ch}$ and $V_{\rm R}$ (with $V_{\rm gu}=0$)

\begin{equation}
N_e=\frac{\varepsilon\varepsilon_0}{ed} \left( S_{\rm ch} V_{\rm ch} + S_{\rm R}V_{\rm R} \right) - \frac{\varepsilon\varepsilon_0}{\alpha e d} S V_{\rm e},
\label{eq:ns}
\end{equation}         

\noindent where $N_{\rm e}$ is the total number of SE in the device, $S_{\rm ch}$ and $S_{\rm R}$ is the area of the charged surface of liquid in the middle segment and reservoirs, respectively, and $S=S_{\rm ch}+S_{\rm R}$. Plugging this relation into (\ref{eq:Eperp1}) with $V_{\rm bot}=V_{\rm ch}$ and $V_{\rm gu}=0$, we obtain a linear dependence between $V_{\rm R}$ and $V_{\rm ch}$ with the slope equal to $(\alpha-1/2)a/[1/2-(\alpha-1/2)b]$, where $a=S_{\rm R}/S$ and $b=S_{\rm ch}/S$. For our device, we have $a=34/40$ and $b=6/40$ (see figure~\ref{fig:2}(a)), which predicts the slope value equal to 0.56. This is in a satisfactory agreement with our experimental result given in figure~\ref{fig:6}(b).

\section{Discussion}
\label{sec:disc}

The experimental results presented in the previous section demonstrate high-level control and tunability of SE in a device comprised of an array of microchanels. The Rydberg energy spectrum of SE in such a device is well described by simple electrostatics, giving a very good agreement with the measured image-charge spectra. This presents SE confined in microchannels as a promising platform for further work towards quantum-state detection and qubit implementation with electrons on helium. In our experiment, an image-charge signal from about $10^5$ electrons excited by the MW radiation is readily observed with the signal-to-noise (SNR) ratio approaching 100, thus showing feasibility to scale down the Rydberg-transition detection to a much smaller number of electrons. The most straightforward approach towards increasing the sensitivity of the image-charge detection is to further increase the change in the image charge $|\delta q|$ for an excited electron, see the discussion in section~\ref{sec:intro}, by employing a microchannel device with a smaller channel depth $d$. However, this would bring electrons closer to the metal electrodes, thus shifting their Rydberg transition frequency into the THz range due to the Stark shift by the electric field of image charges. A more suitable approach is to increase the sensitivity of the image-current detection method employed together with the microchannel setup. Below, we discuss two methods towards such an improvement.  

\begin{figure}[t]
\begin{center}
\includegraphics[width=16cm,keepaspectratio]{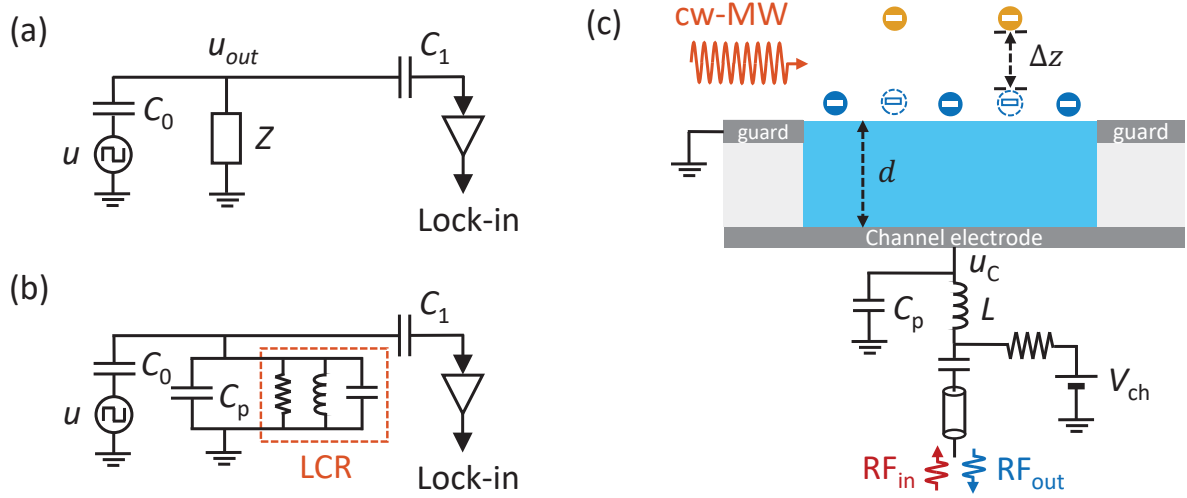}
\end{center}
\caption{\label{fig:7} Thevenin equivalent circuit of the image-charge detection setup with an arbitrary load impedance $Z$ (a) and with a parallel $LCR$ circuit (b). The lock-in amplifier is referenced at the MW modulation frequency $\omega_{\rm m}$ that matches the resonance frequency $\omega_0$ of the tank circuit. (c) RF reflection method for a dispersive readout of the Rydberg excitation of SE in a microchannel. Electrons are excited by continuous wave (cw) microwaves, while their transition probability is modulated at the frequency of the RF driving.}
\end{figure} 

From the equivalent Thevenin circuit of the setup shown in figure~\ref{fig:1}(c), it is clear that the output voltage signal $u_{\rm out}$ is limited by the value of the parasitic capacitance $C_{\rm p}$. In the experiment reported here, this capacitance is about 20~pF and is determined by the capacitance of the connecting cryogenic cable. There are serious technical limitations due to the dilution refrigerator operation which prevent us from shortening this cable, thus  decreasing the value of $C_{\rm p}$. A better solution is to employ a resonant (tank) circuit to eliminate the parasitic capacitance and to increase the output voltage signal. First, let us consider the equivalent Thevenin circuit shown in figure~\ref{fig:6}(a), with a load impedance $Z$. As before, we assume an infinite input impedance of the cryogenic amplifier. Then, the output voltage is given by

\begin{equation}
u_{\rm out} = \frac{Zu}{\left[ Z+(j\omega_{\rm m} C_0)^{-1} \right]}.
\label{eq:Z}
\end{equation}        

\noindent Next, let us consider the load impedance in the form of a parallel $LCR$ circuit, with $Z^{-1}=j\omega_{\rm m} C + R^{-1} + (j\omega_{\rm m} L)^{-1}$. Plugging this into (\ref{eq:Z}) and using $u=-\Delta q/C_0$, we obtain 

\begin{equation}
u_{\rm out} = \frac{-j\Delta q \omega_{\rm m} }{\left[ j(C_0+C)\omega_{\rm m}^{-1}(\omega_{\rm m}^2-\omega_0^2) + R^{-1} \right]},
\label{eq:LCR}
\end{equation}   

\noindent where $\omega_0=[L(C_0+C)]^{-1/2}$. Note that at the resonance condition corresponding to $\omega_{\rm m}=\omega_0$ we obtain $u_{\rm out}=-j R \Delta q \omega_{\rm m} $. In other words, the output voltage signal amplitude is given by $iR$, where $i$ is the equivalent Norton current (see figure~\ref{fig:1}(b)) and $R$ is the impedance of the parallel $LCR$ circuit at resonance. Note that such a circuit eliminates any parasitic capacitance $C_{\rm p}$ (see figure~\ref{fig:6}(b)) which can be simply added to the capacitance of the tank circuit causing a shift in its resonance frequency $\omega_0$.  

For the parallel $LCR$ circuit, a promising approach is to use a lumped-element superconducting resonator~\cite{WinlRRA2017}. Such resonators can be placed at cryogenic temperatures in proximity to the experimental cell. When cooled below the superconducting transition temperature, they achieve a high quality factor $Q\sim 10^4\textrm{--}10^5$ at the resonance frequency $\omega_0/2\pi\sim 1$~MHz, even in the presence of a strong magnetic field~\cite{UlmRSI2009,UlmRSI2016}. This results in a high value of the effective parallel resistance $R=Q \omega_0 L \sim 10^2 \textrm{--}10^3$~M$\Omega$. Even with limitations associated with a finite impedance of the cryogenic amplifier and added losses in the circuit, this method allows detection of the image current on the order of 10~fA~\cite{JeffRSI1993}. In the experimental setup reported here, excitation of a single electron will produce an image current $i=\kappa e\omega_{\rm m}(\Delta z/d)\sim 10^{-14}$~A for the MW modulation frequency $\omega_{\rm m}/(2\pi)=1$~MHz. This approach shows the feasibility of detecting a single electron in our setup.       
     
Another approach towards improving the image-charge detection method is based on a dispersive readout of the Rydberg transition, as was outlined previously~\cite{KawaPRL2019}. In this method, we measure a change in the capacitance due to variation of the image charge induced in the capacitor plates by SE excited by the continuous-wave (cw) MW radiation. In order to realize this method, we connect the capacitor $C_0$ formed by the device electrodes in series with an inductor $L$ and drive this tank circuit with an RF signal at the frequency $\omega_{\rm RF}$ close to the resonance frequency of the circuit, see figure~\ref{fig:7}(c). The ac component of the voltage $u_C$ across the capacitor causes a periodic modulation (with the frequency $\omega_{\rm RF}$) of the excited-state population $\rho_{22}$ of SE via the Landau-Zener-St\"{u}ckelberg (LZS) transitions~\cite{NoriPhysRep2010}, providing that the Rydberg transition frequency of SE is sufficiently close to the frequency of the applied MW radiation. This causes a change in the capacitance value given by

\begin{equation}
\Delta C = \frac{q(t)}{u_C(t)} = \frac{\bar{\rho}_{22} N_e\kappa e \Delta z}{\bar{u}_C d},
\label{eq:LCR}
\end{equation}   

\noindent where $\bar{u}_C$ and $\bar{\rho}_{22}$ is the amplitude of the harmonic variation of $u_C$ and $\rho_{22}$, respectively. In order to obtain the maximum $\bar{\rho}_{22}$ by the constructive LZS interference~\cite{NoriPhysRep2010}, SE must be driven through the Rydberg resonance with an amplitude $\bar{u}_C\approx k_1 \Delta \omega_{21}/(2\pi)$, where $k_1$ is the slope of the linear plot in Fig.~\ref{fig:4}(a) and $\Delta \omega_{21}$ is the broadening of the transition line. Using results of the experiment reported here, we estimate the change in capacitance $\Delta C\sim 10^{-16}$~F for $N_e=10^5$, $\Delta \omega_{21}=10$~GHz, and by assuming $\bar{\rho}_{22}=0.1$~\cite{remark}. Note that for a single electron the inhomogeneous broadening of the transition line must be replaced with the intrinsic linewidth due to the ripplon scattering. For sufficiently low temperatures below 100~mK, the intrinsic linewidth is expected to be on the order of 1~MHz~\cite{LeaPRL2002}. This corresponds to the capacitance change of $10$~aF for a single-electron excitation, which can be detected by observing the spectral shift in the RF signal reflected from the tank circuit~\cite{GonzNatCommun2015}, see figure~\ref{fig:7}(c). This makes such a dispersive readout of the single-electron Rydberg transition to be very promising. Note that in this method any parasitic capacitance $C_{\rm p}$ also contributes to the tank circuit (see figure~\ref{fig:7}(c)), therefore to the dispersive changes in the circuit. Thus, the value of this capacitance has to be minimized. In practice, this value is limited to about 1~pF~\cite{GonzNatCommun2015}.                
     
\section{Conclusion}
\label{sec:conc}

We reported on the first observation of the Rydberg transition of SE on superfluid helium in a microchanel device by the image-charge detection method. This work is largely motivated by the possibility for the quantum-state detection of a single electron in a nano-structured device for qubit implementation. It is shown that the sensitivity of the image-charge detection method enhances thousandfold comparing with the previous work~\cite{KawaPRL2019} due to close proximity of SE from the electrodes in the microchannel structure. The observed Rydberg energy spectrum can be controlled with high accuracy by the dc bias voltages applied to the device and is in a very good agreement with our theoretical calculations. It is shown that, by improving the image-current measurement circuit, this detection method can be scaled down to the detection of the Rydberg transition of a single confined electron. This would open a pathway towards the spin-state detection, which has not been demonstrated in this system yet. Together with the high-level control of the electron transport, which was demonstrated in the microchannel devices previously~\cite{GlasPRL2001,IkegPRL2009,ReesPRL2016,BadrPRL2020,ReesPRL2011,ZouJLTP2022}, our results show that SE confined in such devices present a suitable platform for electron manipulation and their quantum-state detection, with further prospects for qubit implementation and building a scalable quantum computer~\cite{BradPRL2011,YuPRA2022}.            

\ack

This work is supported by the internal grant "KICKS" from Okinawa Institute of Science and Technology (OIST) Graduate University. We thank Erika Kawakami for helpful discussions and Research Support Division of OIST Graduate University for technical support.


\begin{thebibliography}{99}
\bibitem{remark} Details of the theoretical calculations for the excited-state population of SE due the Landau-Zener-Stuckelberg interference will be presented elsewhere.
\end{thebibliography}
\end{document}